\documentclass[
 aip,
 amsmath,amssymb,
 twocolumn,
]{revtex4}

\usepackage{textcase}
\usepackage{graphicx}% Include figure files
\usepackage{dcolumn}% Align table columns on decimal point
\usepackage{bm}% bold math

\usepackage{amsmath,amsthm,amssymb,mathrsfs}
\usepackage{bm}
\usepackage{subfigure}
\newcommand{\etal}{\textit{et\ al.\ }}

\newcommand{\D}{{\rm d}}
\newcommand{\E}{{\rm e}}

\newcommand{\Plog}{{\rm plog}}
\newcommand{\Erf}{{\rm Erf}}

\begin{document}

%====================================================================================================================================================%

\title{Dependence of spin torque switching
probability on electric current}

\author{Tomohiro Taniguchi${}^{1,2}$}
\author{Hiroshi Imamura${}^{2}$}
\email{h-imamura@aist.go.jp}

\affiliation{ 
$^{1}$Magnetic Materials Unit,
National Institute for Materials Science, 1-2-1 Sengen, Tsukuba 305-0047, Japan \\
$^{2}$Nanoscale Theory Group, Nanosystem Research Institute, 
      National Institute of Advanced Industrial Science and Technology, 1-1-1 Umezono, Tsukuba 305-8568, Japan
}

\date{\today}% 

%====================================================================================================================================================%

\begin{abstract}
  Dependence of the thermally assisted spin torque switching probability 
  on the sweep electric current 
  was investigated theoretically. 
  The analytical expressions of 
  the switching times for $b=1$ and $b=2$ are derived 
  based on the rate equation, 
  where $b$ is the exponent of the current term in the switching rate. 
  The switching current is approximately proportional to 
  the temperature $T$ and the logarithm of the sweep rate $v$ 
  for both $b=1$ and $b=2$ 
  in the experimentally performed ranges of $T$ and $v$. 
  Experiments in very low temperature range are required 
  to determine the exponent $b$. 
\end{abstract}

%\keywords{Spin}%Use showkeys class option if keyword
\maketitle

%====================================================================================================================================================%

\section{Introduction}
\label{sec:Introduction}

Spin random access memory (Spin RAM), 
which employs
spin torque switching \cite{slonczewski96,berger96} 
as the writing method, 
is one of the key spin-electronics applications 
for future nanotechnology. 
In Spin RAM application,
a high thermal stability $\Delta_{0}$ of 
more than 40 is required 
to guarantee retention time longer than 10 years. 
Recently, Hayakawa \etal and Yakata \etal studied 
the thermal stabilities of 
antiferromagnetically and ferromagnetically coupled synthetic free layers 
respectively, 
and found higher thermal stabilities 
compared with that of a single free layer
\cite{hayakawa08,yakata09,yakata10}. 

%====================================================================================================================================================%

The thermal stability is determined by 
the thermally assisted magnetization switching probability 
under the effect of 
the applied magnetic field or spin torque. 
The magnetic field or spin torque is 
assumed to be small compared to 
the anisotropy field $H_{\rm K}$ or critical current $I_{\rm c}$ at $T=0$. 
In 1963
Brown derived the analytical expression of the switching probability 
which includes the effect of constant field $H_{\rm appl}$ \cite{brown63} 
which is given by
$P=1-\exp\{-f_{0}t\exp[-\Delta_{0}(1-H_{\rm appl}/H_{\rm K})^{2}]\}$, 
where $f_{0}$ is the attempt frequency. 
The switching probability under the effect of 
constant current $I$ (spin torque) was independently derived 
by Koch \etal \cite{koch04} and Li and Zhang \cite{li04} in 2004, 
and is given by 
$P=1-\exp\{-f_{0}t\exp[-\Delta_{0}(1-H_{\rm appl}/H_{\rm K})^{2}(1-I/I_{\rm c})]\}$. 
It should be noted that 
the exponent of the current term, $(1-I/I_{\rm c})$, 
in this formula is unity. 
On the other hand, 
Suzuki \etal showed that the switching probability is given by 
$P=1-\exp\{-f_{0}t\exp[-\Delta_{0}(1-H_{\rm appl}/H_{\rm K}-I/I_{\rm c})^{2}]\}$
\cite{suzuki09}, 
where the exponent of the current term is $2$. 
Recently, 
we derived the switching probability formula which is 
given by $P=1-\exp\{-f_{0}t\exp[-\Delta_{0}(1-H_{\rm appl}/H_{\rm K})^{2}(1-I/I_{\rm c})^{2}]\}$
\cite{taniguchi11}, 
which slightly different compared to the result of Suzuki \etal 
but agrees well with the results of Butler \etal \cite{comment1}. 
The determination of the value of 
the exponent $b$ of the current term 
in the switching probability formula, 
$P=1-\exp\{-f_{0}t\exp[-\Delta_{0}(1-H_{\rm appl}/H_{\rm K})^{2}(1-I/I_{\rm c})^{b}]\}$,
remains unsettled, 
although it is very important 
for the determination of the thermal stability \cite{taniguchi11}. 

%====================================================================================================================================================%

Recently we \cite{taniguchi11}
pointed out the reason why 
the values of the exponent $b$ for the spin torque switching are 
different between Ref. \cite{koch04} and Ref. \cite{taniguchi11}. 
In the case of the completely in-plane switching of 
the in-plane magnetized system, 
or in the case of the perpendicularly magnetized system, 
the switching can be described by only the angle from the easy axis, 
and the effect of spin torque can be 
regarded as the enhancement of the applied field, 
as shown in Ref. \cite{suzuki09,taniguchi11,comment1}. 
Thus, the exponent $b$ is reduced to $2$, 
as is in the case for the field switching \cite{brown63}. 
However, the results of Refs. \cite{koch04,li04} ($b=1$) 
are supported by the experimental result \cite{myers02}, 
in which the mean switching current 
is approximately proportional to 
the temperature $T$ 
and the logarithm of the current sweep rate $\log v$. 

%====================================================================================================================================================%

In this paper, 
we derived the analytical expressions of 
the switching times $t_{\rm sw}$
for $b=1$ and $b=2$ 
due to the time dependent current $I(t)$, 
and calculated the dependence of the switching current $I(t_{\rm sw})$ 
on the temperature $T$ and the sweep rate $v$. 
We found that 
$I(t_{\rm sw})$ is approximately proportional to $T$ and $\log v$ 
for both $b=1$ and $b=2$ 
in the experimental ranges, 
and thus, 
the value of the exponent cannot be determined 
by the results of Ref. \cite{myers02}. 
We also found that 
a low temperature experiment will clarify 
the value of the exponent $b$. 
$I(t_{\rm sw})$ is proportional to $T$ 
if $b=1$ 
while it is nonlinear to $T$ 
if $b=2$. 

%====================================================================================================================================================%

This paper is organized as follows. 
The analytical expressions of the switching probability $P$ 
and the switching time $t_{\rm sw}$ 
for $b=1$ and $b=2$ are derived 
in Sec. \ref{sec:Switching_probability} and 
in Sec. \ref{sec:Switching_time}, respectively. 
The dependences of the switching current $I(t_{\rm sw})$ 
on the temperature $T$ and the sweep rate $v$ are 
discussed in 
Sec. \ref{sec:Dependence_of_switching_time_on_temperature_and_sweep_rate}. 
Section \ref{sec:Summary} is the summary of this paper. 

%====================================================================================================================================================%

\section{Switching probability}
\label{sec:Switching_probability}

In this section, 
we derive the analytical expression of 
the magnetization switching probability 
due to the sweep current 
($b=1$ and $b=2$). 
We consider spin torque switching 
of the free layer magnetization 
in a ferromagnetic (fixed layer)/nonmagnetic/ferromagnetic (free layer) trilayer. 
Both ferromagnetic layers are assumed to 
have the perpendicular anisotropy along the $z$ axis. 
The magnetization $\mathbf{M}$ of the free layer is assumed to 
point to the positive $z$ direction at $t=0$. 
Throughout this paper, 
the applied field $H_{\rm appl}$ is assumed to be zero
for simplicity.
From $t=0$, 
the current $I$ is applied 
to the ferromagnetic film 
perpendicular to plane 
and induces torque on the magnetization $\mathbf{M}$. 
The magnitude of the current $I$ is assumed to be 
smaller than the critical current $I_{\rm c}$. 
With the help of the thermal activation, 
the magnetization can change its direction to 
the negative $z$ direction. 
The switching probability of the magnetization, 
$P$, obeys 
the following equation \cite{brown63}:
\begin{equation}
  \frac{\D P(t)}{\D t}
  =
  r(t)
  \left[
    1 
    - 
    P(t)
  \right],
  \label{eq:rate_equation}
\end{equation}
where $r(t)$ is 
the switching rate 
from the initial state ($\mathbf{M} \parallel \mathbf{e}_{z}$) 
to the final state ($\mathbf{M} \parallel -\mathbf{e}_{z}$). 
The explicit form of $r(t)$ is given by 
\begin{equation}
  r(t)
  =
  f_{0}
  \exp
  \left[
    -\Delta_{0}
    \left(
      1
      -
      \frac{I(t)}{I_{\rm c}}
    \right)^{b}
  \right].
  \label{eq:switching_probability}
\end{equation}
Here $b$ is the exponent of the current term $(1-I/I_{\rm c})$. 
Although the attempt frequency depends on 
$I(t)$ in general, 
we approximate the value of $f_{0}$ as 
the attempt frequency at zero current. 
This approximation is applicable 
because the dependence of the switching rate $r(t)$ on 
the current $I(t)$ is mainly determined by 
the exponential term $\exp[-\Delta_{0}(1-I/I_{\rm c})^{b}]$. 
The attempt frequency with zero current is given by 
$f_{0}=\alpha\gamma H_{\rm K}\sqrt{\Delta_{0}/\pi}/(1+\alpha^{2})$, 
where $\alpha$ and $\gamma$ are 
the Gilbert damping constant and 
the gyromagnetic ratio, respectively \cite{brown63}. 

%====================================================================================================================================================%

Let us consider the thermally assisted 
magnetization switching due to 
time dependent current $I(t)$. 
As is in experiments \cite{yakata10,myers02},
we assume that the strength of $I(t)$ increases 
linearly with time, 
i.e., $I(t)=vt$, where $v$ is the sweep rate. 
By integrating Eq. (\ref{eq:rate_equation}) 
and using the initial condition 
$P(0)=0$, 
$P(t)$ is given by 
\begin{equation}
\begin{split}
  P(t)
  =
  1-
  &
  \exp
  \left\{
    -\frac{f_{0}I_{\rm c}}{bv\Delta_{0}^{1/b}}
    \left[
      \gamma
      \left(
        \frac{1}{b},
        \Delta_{0}
      \right)
    \right.
  \right.
\\
  &\ \ \ \ \ \ \ \ \ \ \ \ \ \ \ \ \ \ -
  \left. 
    \left.
      \gamma
      \left(
        \frac{1}{b},
        \Delta_{0}
        \left(
          1
          -
          \frac{I}{I_{\rm c}}
        \right)^{b}
      \right)
    \right]
  \right\},
\end{split}
\end{equation}
where 
$\gamma(\beta,z)=\int_{0}^{z}\D t t^{\beta-1}\E^{-t}$ is 
the incomplete $\Gamma$ function. 
For $b=1$ \cite{koch04,li04} 
and $b=2$ \cite{brown63,suzuki09,taniguchi11,comment1}, 
$P(t)$ are, respectively, reduced to 
\begin{equation}
\begin{split}
  P_{1}(t)
  =
  1-
  &
  \exp
  \left\{
    -\frac{f_{0}I_{\rm c}}{v\Delta_{0}}
    \left[
      \exp
      \left[
        -\Delta_{0}
        \left(
          1
          -
          \frac{vt}{I_{\rm c}}
        \right)
      \right]
    \right.
  \right.
\\
  &\ \ \ \ \ \ \ \ \ \ \ \ \ \ \ \ \ \ \ \ \ -
  \left. 
    \left.
      \exp
      \left(
        -\Delta_{0}
      \right)
    \right]
  \right\},
  \label{eq:R_1}
\end{split}
\end{equation}
\begin{equation}
\begin{split}
  P_{2}(t) 
  =
  1 - 
  &
  \exp
  \left\{
    -\frac{\sqrt{\pi}f_{0}I_{\rm c}}{2v\sqrt{\Delta_{0}}}
    \left[
      \Erf
      \left(
        \sqrt{\Delta_{0}}
      \right)
    \right.
  \right.
\\
  &\ \ \ \ \ \ \ \ \ \ \ \ \ \ \ \ \ \ \ \ \ -
  \left.
    \left.
      \Erf
      \left(
        \sqrt{\Delta_{0}}
        \left(
          1
          -
          \frac{vt}{I_{\rm c}}
        \right)
      \right)
    \right]
  \right\},
  \label{eq:R_2}
\end{split}
\end{equation}
where $\Erf(z)=(2/\sqrt{\pi})\int_{0}^{z}\D t \E^{-t^{2}}$ is the error function.
Equations (\ref{eq:R_1}) and (\ref{eq:R_2}) are 
the main results in this section. 

%====================================================================================================================================================%

\begin{figure}%[p]
\centerline{\includegraphics[width=0.6\columnwidth]{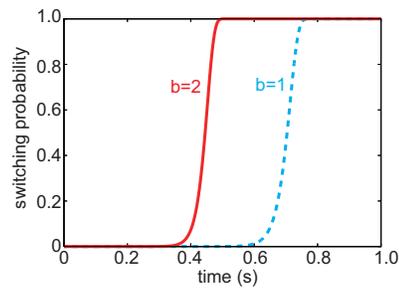}}
\caption{
         The time evolutions of $P(t)$ 
         for $b=1$ ($P_{1}$ in Eq. (\ref{eq:R_1})) 
         and $b=2$ ($P_{2}$ in Eq. (\ref{eq:R_2})). 
         \vspace{-3ex}}
\label{fig:fig1}
\end{figure}

%====================================================================================================================================================%

Figure \ref{fig:fig1} shows 
the time evolutions of $P_{1}$ and $P_{2}$ 
under the effect of the sweep current. 
The values of the parameters are taken to be 
$\alpha=0.007$, $\gamma=1.764\times 10^{7}$ Hz/Oe, 
$H_{\rm K}=200$ Oe, 
$M=995$ emu/c.c.,
$T=300$ K,
$d=2.0$ nm and 
$S=\pi\times 80\times 35$ nm${}^{2}$,
where $d$ and $S$ are 
the thickness and cross section area 
of the ferromagnetic layer, respectively, 
which are typical values for the Spin RAM cell 
consists of the transition ferromagnetic metals 
such as CoFeB 
\cite{yakata09,yakata10}. 
The thermal stability is defined by 
$\Delta_{0}=MH_{\rm K}V/(2k_{\rm B}T)$, 
where $V=Sd$ is the volume. 
The critical current $I_{\rm c}$ is given by 
$I_{\rm c}=[2\alpha eMSd/(\hbar g)](H_{\rm K}+2\pi M)$ \cite{sun00},
where spin polarization $g$ is taken to be $0.5$. 
By using the above parameters, 
the values of $f_{0}$, $\Delta_{0}$, and $I_{\rm c}/S$ are 
given by 
$0.09$ GHz, $42$, and $5.5\times 10^{6}$ A/cm${}^{2}$, 
respectively. 
The sweep rate is taken to be 
$v/S=5.0\times 10^{6}$ A/cm${}^{2}$ s 
($v=0.44$ mA/s), 
which is similar to the experimental values \cite{myers02}. 

%====================================================================================================================================================%

The switching probability $P$ 
suddenly changes its value 
from 0 to 1 
at a certain time $t_{\rm sw}$, 
as shown in Fig. \ref{fig:fig1}, 
in which $t_{\rm sw}$ satisfies 
$I(t_{\rm sw})<I_{\rm c}$. 
In next section, 
we derive the analytical expression of 
the switching time $t_{\rm sw}$.

%====================================================================================================================================================%

%====================================================================================================================================================%

\section{Switching time}
\label{sec:Switching_time}

In this section, 
we derive the analytical expression of 
the switching time. 
At $t=t_{\rm sw}$, 
$\D P/\D t$ takes 
its maximum, as shown in Fig. \ref{fig:fig1}. 
Thus, $P(t)$ satisfies 
$\D^{2}P/\D t^{2}=0$ at $t=t_{\rm sw}$. 
This condition can be rewritten as 
\begin{equation}
  \frac{\D r}{\D t}
  =
  r^{2}.
\end{equation}
By using the explicit form of $r(t)$ (Eq. (\ref{eq:switching_probability})), 
we find that 
\begin{equation}
  \frac{bv\Delta_{0}}{I_{\rm c}}
  \left(
    1
    -
    \frac{I(t_{\rm sw})}{I_{\rm c}}
  \right)^{b-1}
  =
  f_{0}
  \exp
  \left[
    -\Delta_{0}
    \left(
      1
      -
      \frac{I(t_{\rm sw})}{I_{\rm c}}
    \right)^{b}
  \right].
  \label{eq:transition_time_equation}
\end{equation}
In the case of $b=1$, 
Eq. (\ref{eq:transition_time_equation}) can be easily solved, 
and we find that 
\begin{equation}
  \frac{I(t_{\rm sw})}{I_{\rm c}}
  =
  1
  -
  \frac{1}{\Delta_{0}}
  \log 
  \frac{f_{0}I_{\rm c}}{v\Delta_{0}}. 
  \label{eq:F_Fc_1}
\end{equation}
By using $I=vt$, 
the switching time for $b=1$ is given by 
\begin{equation}
  t_{\rm sw1}
  =
  \frac{I_{\rm c}}{v}
  \left(
    1
    -
    \frac{1}{\Delta_{0}}
    \log
    \frac{f_{0}I_{\rm c}}{v\Delta_{0}}
  \right),
  \label{eq:switching_time_1}
\end{equation}
which is identical to 
that obtained by Li and Zhang \cite{li04} 
for $P=1-\E^{-1}$ 
(see Eq. (22) in Ref. \cite{li04}). 

%====================================================================================================================================================%

To obtain the explicit form of $t_{\rm sw}$ for general $b$, 
with the help of Eq. (\ref{eq:F_Fc_1}), 
let us consider the solution of $I(t_{\rm sw})/I_{\rm c}$ 
whose form is given by 
\begin{equation}
  \frac{I(t_{\rm sw})}{I_{\rm c}}
  =
  1
  -
  \left(
    \frac{1}{\Delta_{0}}
    \log X
  \right)^{1/b},
  \label{eq:assumed_F_Fc_b}
\end{equation}
where $X$ is determined by Eq. (\ref{eq:transition_time_equation}). 
By substituting Eq. (\ref{eq:assumed_F_Fc_b}) into Eq. (\ref{eq:transition_time_equation}), 
we find that $X$ satisfies 
$\left( \log X \right)^{1-1/b}=1/(CX)$, 
where $C=bv\Delta_{0}^{1/b}/(f_{0}I_{\rm c})$. 
In terms of $Y=\log X$, 
this equation can be expressed as 
$Y\E^{Y}Y^{-1/b}=1/C$, 
and its solution is given by 
$Y=[(b-1)/b]\Plog[bC^{-b/(b-1)}/(b-1)]$, 
where $\Plog(z)$ is the product log 
(or Lambert $W$ function) 
which satisfies 
$\Plog(z)\exp[\Plog(z)]=z$. 
$I(t_{\rm sw})/I_{\rm c}$ is thus given by 
\begin{equation}
  \frac{I(t_{\rm sw})}{I_{\rm c}}
  =
  1
  -
  \left\{
    \frac{b-1}{b\Delta_{0}}
    \Plog
    \left[
      \frac{b}{b-1}
      \left(
        \frac{f_{0}I_{\rm c}}{bv\Delta_{0}^{1/b}}
      \right)^{b/(b-1)}
    \right]
  \right\}^{1/b}.
  \label{eq:F_Fc_b}
\end{equation}
In the case of $b=2$, 
the switching time $t_{\rm sw2}$ is given by 
\begin{equation}
  t_{\rm sw2}
  =
  \frac{I_{\rm c}}{v}
  \left\{
    1
    -
    \sqrt{
      \frac{1}{2\Delta_{0}}
      \Plog
      \left[
        \left(
          \frac{f_{0}I_{\rm c}}{\sqrt{2}v\sqrt{\Delta_{0}}}
        \right)^{2}
      \right]
    }
  \right\}.
  \label{eq:switching_time_2}
\end{equation}
Equations (\ref{eq:switching_time_1}) and (\ref{eq:switching_time_2}) are
the main results in this section: 
these are the theoretical expressions of 
the switching times $t_{\rm sw}$,
or equivalently, 
the switching current $I(t_{\rm sw})=vt_{\rm sw}$. 
They are valid for the sweep rate $v$ 
which satisfies $t_{\rm sw}>0$. 
By using the parameter values 
listed in the previous section, 
$t_{\rm sw1}$ and $t_{\rm sw2}$ are estimated to be 
$0.71$ s and $0.45$ s, respectively, 
which have good agreement 
with Fig. \ref{fig:fig1}. 

%====================================================================================================================================================%

\begin{figure}%[p]
\centerline{
\includegraphics[width=1.0\columnwidth]{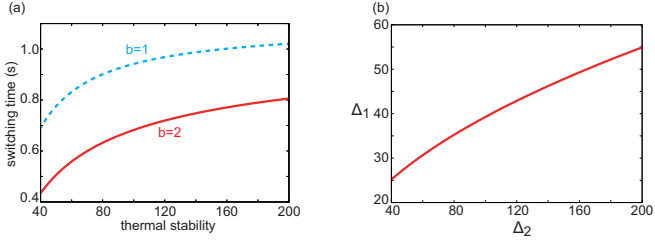}}
\caption{
         (a) The dependences of the switching times 
         for $b=1$ and $b=2$ 
         (Eqs. (\ref{eq:switching_time_1}) and (\ref{eq:switching_time_2})) 
         on the thermal stability $\Delta_{0}$.
         (b) The relation between the thermal stabilities 
         estimated by Eqs. (\ref{eq:switching_time_1}) and (\ref{eq:switching_time_2}). 
\vspace{-3ex}}
\label{fig:fig2}
\end{figure}

%====================================================================================================================================================%

For a large $z$, $\Plog(z)$ can be 
approximated to $\log(z)$. 
In this limit, 
Eq. (\ref{eq:F_Fc_b}) is reduced to 
\begin{equation}
  \frac{I(t_{\rm sw})}{I_{\rm c}}
  \simeq 
  1
  -
  \left\{
    \frac{1}{\Delta_{0}}
    \log 
    \left[
      \left(
        \frac{b}{b-1}
      \right)^{1-1/b}
      \frac{f_{0}I_{\rm c}}{bv\Delta_{0}^{1/b}}
    \right]
  \right\}^{1/b}. 
  \label{eq:F_Fc_b_approx}
\end{equation}
Neglecting a small correlation term $[b/(b-1)]^{1-1/b}$, 
Eq. (\ref{eq:F_Fc_b_approx}) is identical to 
the result of Garg \cite{garg95} 
on the mean switching current 
$\langle I \rangle/I_{\rm c}=1-\{(1/\Delta_{0})\log[f_{0}I_{\rm c}/(bv\Delta_{0}^{1/b})]\}^{1/b}$ 
which was used in the analysis of the experimental result of Ref. \cite{myers02}. 

%====================================================================================================================================================%

Experimentalists \cite{yakata10,myers02} repeat 
the measurement of the switching 
due to sweep current or magnetic field
and measure the switching time 
of each trial. 
By fitting the average of the switching time 
with Eqs. (\ref{eq:switching_time_1}) or (\ref{eq:switching_time_2}), 
one can estimate the thermal stability $\Delta_{0}$. 
Figure \ref{fig:fig2} (a) shows 
the dependence of the switching times, 
Eqs. (\ref{eq:switching_time_1}) and (\ref{eq:switching_time_2}), 
on the thermal stability $\Delta_{0}$, 
in which $\Delta_{0}$ is regarded as 
the independent variable from $I_{\rm c}$, 
as is in experiments \cite{yakata09,yakata10}. 
For example, 
when the switching time is about 0.8 s, 
the thermal stability estimated by Eq. (\ref{eq:switching_time_1}) is about 50 
while that estimated by Eq. (\ref{eq:switching_time_2}) is about 200. 
Figure \ref{fig:fig2} (b) shows 
the relation between the thermal stabilities 
estimated by Eqs. (\ref{eq:switching_time_1}) ($\Delta_{1}$) 
and (\ref{eq:switching_time_2}) ($\Delta_{2}$)
which can be obtained by putting $t_{\rm sw1}(\Delta_{0}=\Delta_{1})=t_{\rm sw2}(\Delta_{0}=\Delta_{2})$ 
and is given by 
\begin{equation}
\begin{split}
  \Delta_{1}
  &=
  \sqrt{2\Delta_{2}}
  \Plog
  \left\{
    \frac{f_{0}I_{\rm c}}{\sqrt{2}v\sqrt{\Delta_{2}}}
    \sqrt{
      \Plog
      \left[
        \left(
          \frac{f_{0}I_{\rm c}}{\sqrt{2}v\sqrt{\Delta_{2}}}
        \right)^{2}
      \right]
    }
  \right\}
\\
  &\ \ \ \ \ \ \ \ \ \ \bigg/
  \sqrt{
    \Plog
    \left[
      \left(
        \frac{f_{0}I_{\rm c}}{\sqrt{2}v\sqrt{\Delta_{2}}}
      \right)^{2}
    \right]
  }
\\
  &\simeq 
  \sqrt{\Delta_{2}}
  \sqrt{
    \log \frac{f_{0}I_{\rm c}}{\sqrt{2}v\sqrt{\Delta_{2}}}
  }.  
  \label{eq:delta_1_delta_2}
\end{split}
\end{equation}
As shown in Fig. \ref{fig:fig2} (b),
the difference of the exponent $b$ 
leads to a significant underestimation 
of the thermal stability. 

%====================================================================================================================================================%

\section{Dependence of switching time on temperature and sweep rate}
\label{sec:Dependence_of_switching_time_on_temperature_and_sweep_rate}

Myers \etal experimentally studied 
the dependences of the switching current, 
$I(t_{\rm sw})$, 
on the temperature $T$ and the sweep rate $v$ 
(see Fig. 1 (d) in Ref. \cite{myers02}), 
and showed that $I(t_{\rm sw})$ is approximately 
proportional to both $T$ and $\log v$. 
Reference \cite{li04} argued that 
these results support the theory with $b=1$ \cite{koch04,li04}
because such dependences 
are expected from Eq. (\ref{eq:F_Fc_1}). 
In this section, 
we study the dependences of $I(t_{\rm sw})$ for $b=1$ and $b=2$ 
on $T$ and $v$ 
in and out of the experimental ranges. 

%====================================================================================================================================================%

%====================================================================================================================================================%

\begin{figure}%[p]
\centerline{
\includegraphics[width=0.6\columnwidth]{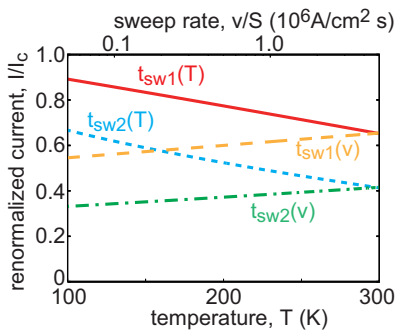}}
\caption{
         The solid (red) and dotted (blue) lines are 
         the dependences of $I(t_{\rm sw})/I_{\rm c}$ 
         on the temperature $T$ for $b=1$ and $b=2$, respectively. 
         The dashed (orange) and dashed-dotted lines are 
         the dependences of $I(t_{\rm sw})/I_{\rm c}$ 
         on the sweep rate $v$ for $b=1$ and $b=2$, respectively. 
\vspace{-3ex}}
\label{fig:fig3}
\end{figure}

%====================================================================================================================================================%

%====================================================================================================================================================%

The dependence of $I(t_{\rm sw})$ on the temperature 
can be taken into account 
through the thermal stability $\Delta_{0}=MH_{\rm K}V/(2k_{\rm B}T)$. 
For simplicity, 
we neglect the dependence of the attempt frequency 
$f_{0} \propto \sqrt{\Delta_{0}}$ 
on the temperature $T$, 
as done in Ref. \cite{li04}. 
The solid (red) and dotted (blue) lines 
in Fig. \ref{fig:fig3} show 
the dependences of $I(t_{\rm sw})/I_{\rm c}$ 
on $T$ for $b=1$ and $b=2$, respectively. 
The temperature range (100-300 K) is sufficient 
to consider the $I(t_{\rm sw})$ 
in the experimental range (180-220 K in Ref. \cite{myers02}). 
Even in such wide range of the temperature, 
both $I(t_{\rm sw})/I_{\rm c}$ 
for $b=1$ and $b=2$ are approximately 
proportional to the temperature $T$. 
The dependences of $I(t_{\rm sw})/I_{\rm c}$ on $v$ 
for $b=1$ and $b=2$ are also plotted 
by the dashed (orange) and dashed-dotted (green) lines 
in Fig. \ref{fig:fig3}, respectively. 
The temperature is fixed to $T=300$ K. 
The range of $v/S$, 
from 0.05 to 5.0 $\times 10^{6}$ A/cm${}^{2}$ s 
(from 0.0044 to 0.44 mA/s), 
is very similar to 
that studied in experiments 
(from 0.01 to 1.0 mA/s). 
As shown in Fig. \ref{fig:fig3}, 
both $I(t_{\rm sw})/I_{\rm c}$ for $b=1$ and $b=2$ are 
approximately proportional to $\log v$. 
Summarizing the above results, 
the value of the exponent $b$ ($b=1$ or $b=2$) 
cannot be determined 
by the dependences of $I(t_{\rm sw})/I_{\rm c}$ 
on the temperature $T$ and the sweep rate $v$ 
in the experimental range.

%====================================================================================================================================================%

%====================================================================================================================================================%

\begin{figure}%[p]
\centerline{
\includegraphics[width=1.0\columnwidth]{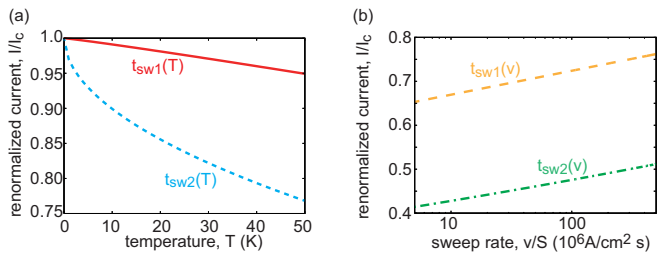}}
\caption{
         The dependences of $I(t_{\rm sw})/I_{\rm c}$ 
         on (a) temperature $T$ and (b) the sweep rate $v$
         out of the experimental range of Ref. \cite{myers02}. 
\vspace{-3ex}}
\label{fig:fig4}
\end{figure}

%====================================================================================================================================================%

%====================================================================================================================================================%

We also study the dependences of $I(t_{\rm sw})/I_{\rm c}$ 
on $T$ and $\log v$ out of the experimental ranges of Ref. \cite{myers02}. 
Figure \ref{fig:fig4} (a) shows the dependence of $I(t_{\rm sw})/I_{\rm c}$ 
on $T$ in the low temperature region, 
$0 < T \le 50$ K. 
In this range, we can see that 
$I(t_{\rm sw})/I_{\rm c}$ is approximately proportional to $T$ 
for $b=1$ 
while $I(t_{\rm sw})/I_{\rm c}$ for $b=2$ is not. 
Thus, to determine the exponent $b$, 
experiments should be performed in such very low temperature region. 
The dependences of $I(t_{\rm sw})/I_{\rm c}$ on $\log v$ 
for both $b=1$ and $b=2$ are approximately proportional to $\log v$ 
even in the large sweep rate region up to 
$v/S=500 \times 10^{6}$ A/cm${}^{2}$ s, 
as shown in Fig. \ref{fig:fig4} (b), 
and thus, 
it is difficult to determine $b$ from 
these dependences. 

%====================================================================================================================================================%

\section{Summary}
\label{sec:Summary}

In summary, 
we studied the dependence of the spin torque switching probability 
on the sweep current. 
The analytical expressions of the switching time 
for both $b=1$ and $b=2$ are derived, 
where $b$ is the exponent of the current term 
in the switching rate. 
We showed that $I(t_{\rm sw})/I_{\rm c}$ 
for $b=1$ and $b=2$ are approximately proportional to 
the temperature $T$ and the logarithm of the sweep rate $v$ 
in the experimental ranges of $T$ and $v$, 
and thus, the value of the exponent $b$ cannot be determined 
by the dependences of $I(t_{\rm sw})/I_{\rm c}$ 
on $T$ and $\log v$. 
We also showed that 
a low temperature experiment is required 
to determine the exponent $b$. 
$I(t_{\rm sw})$ is proportional to $T$ 
if $b=1$ 
while it is nonlinear to $T$ 
if $b=2$.

%====================================================================================================================================================%

\section*{Acknowledgments}

The authors would like to acknowledge 
S. Yuasa, 
H. Kubota, 
S. Yakata, 
D. Bang, 
T. Saruya, 
T. Yorozu, 
K. Seki, 
M. Marthaler, 
and H. Sukegawa 
for the valuable discussions they had with. 
This work was supported by JSPS and NEDO.

%====================================================================================================================================================%

%====================================================================================================================================================%

%\nocite{*}

%\bibliography{biblist}% Produces the bibliography via BibTeX.

\begin{thebibliography}{14}
\bibitem[{com()}]{comment1}
\bibinfo{note}{Private communication with C. Mewes. Their results were recently
  presented at 55${}^{\rm th}$ Annual Conference on Magnetism and Magnetic
  Materials by W. Butler (presentation number is HC-09).}
\end{thebibliography}

%====================================================================================================================================================%

\end{document}